# COMPUTATIONAL FLUID DYNAMICS MODELING OF A WOOD-BURNING STOVE-HEATED TRADITIONAL SAUNA USING NIST'S FIRE DYNAMICS SIMULATOR

Corentin Macqueron, Computational Fluid Dynamics engineer, 2014
corentin.macqueron@gmail.com

**Abstract**

The traditional sauna is studied from a thermal and fluid dynamics standpoint using the NIST's Fire Dynamics Simulator (FDS) software. Calculations are performed in order to determine temperature and velocity fields, heat flux, soot and steam cloud transport, etc. Results are discussed in order to assess the reliability of this new kind of utilization of the FDS fire safety engineering software.

**Keywords**

*Computational Fluid Dynamics, CFD, sauna, wood, Fire Dynamics Simulator, FDS, fluid dynamics, thermo-aeraulics, steam cloud, modeling, combustion*

**Introduction**

The traditional sauna is a wood-burning stove-heated insulated room built for dry bath activities. Besides its physiological effects [9] [10], many physical phenomena occurring inside a sauna are worth of scientific inquiry. Apart from reference [1] (Fan, Holmberg and Heikkinen, 1994), few detailed thermal studies have been performed on the matter.

Recent progresses in fire modeling [2] and rapidly increasing computational resources have made possible the development of Computational Fluid Dynamics (CFD) models of such dry bath rooms.

This article aims to present what can now be done through modeling in order to study the traditional sauna from a thermal and fluid dynamics engineering standpoint, using the NIST's Fire Dynamics Simulator (FDS) software.

**FDS**

FDS is a CFD software developed by the NIST [3] [4] [5] [8]. It is a fast and robust tool to perform numerical fire simulations in the framework of fire safety engineering. It has been validated for many fire engineering applications by comparisons with experimentations [6], but wood-burning stove-heated sauna modeling is a new application presented in this article.

**Presentation of the study**

The sauna studied in this article is a small room (~8 m$^3$) built with wood panels and insulating materials, shown on Figure 1.

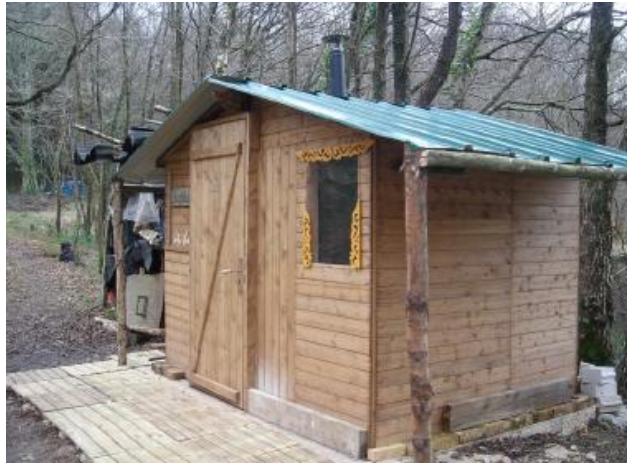

**Figure 1 – Sauna**

The sauna is heated with a small wood-burning stove, shown on Figure 2.

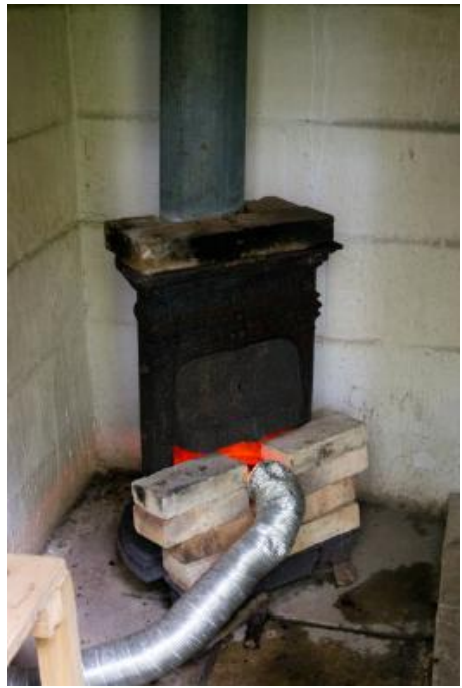

**Figure 2 – Wood-burning stove**

There are no precise temperature, heat-flux or mass flow rate measurements device in the sauna, but a commercial sauna thermometer is installed on one of the internal walls.

The sauna is modeled in FDS as shown on Figure 3. The mesh (computational grid) is shown on Figure 4 and is made of around 170 000 hexahedral cubic cells (4 cm width).

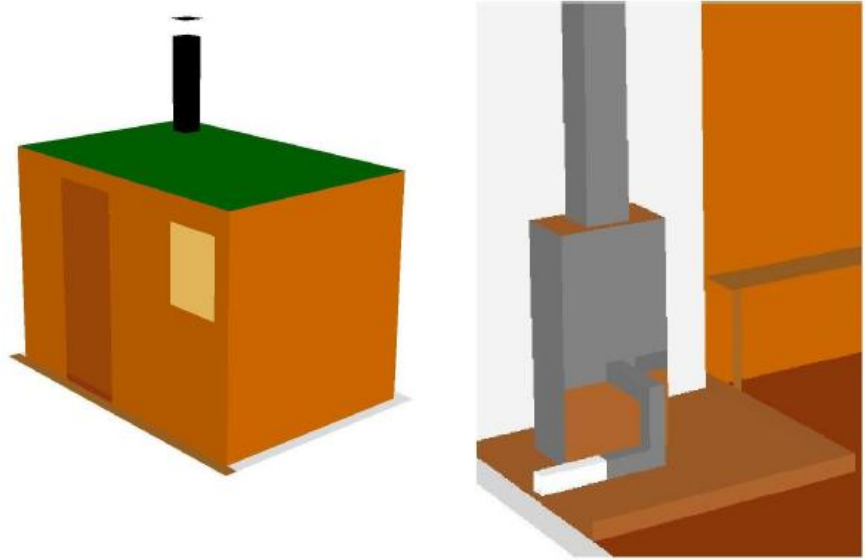

**Figure 3 – FDS model (complete sauna on the left, stove on the right)**

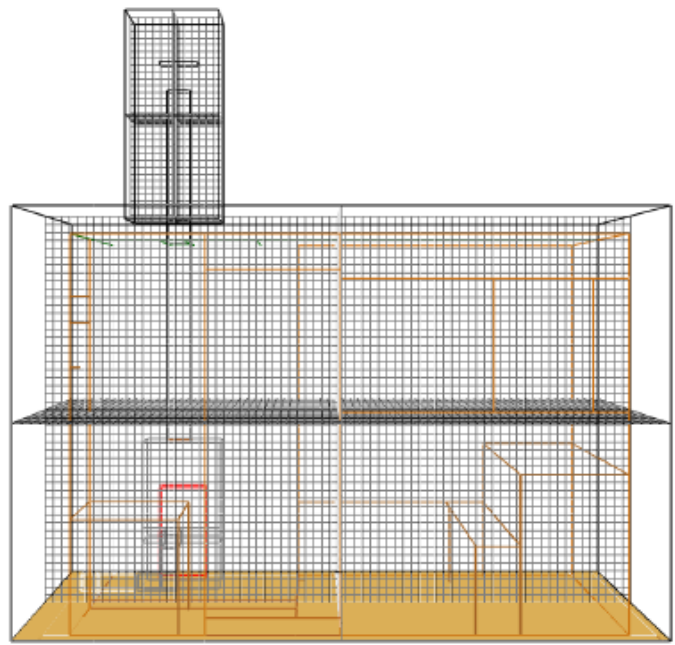

**Figure 4 – Mesh**

One of the main parameter in this type of calculation is the heat source from the fire. As discussed in reference [7], the wood combustion is a very difficult phenomenon to model. The chemical reaction for the wood burning, the soot production and the heat of combustion are taken from reference [11]. The heat release rate value has simply been adjusted in order to reproduce the temperature commonly indicated by the thermometer in steady-state, which is 80°C. The prescribed thermal power required to obtain this temperature is 8.75 kW, which is consistent for an 8 m$^3$ sauna, according to sauna wood stove producers prescriptions [20] and with a previous study of the airflow in a similar sauna [1].

**Main results**

The air temperature is around 100°C, which is what to be expected in a sauna and is quite homogeneous except for the bottom part (due to cold air leaking under the door), as shown on Figure 5. The gases in the chimney, at the exhaust of the stove, are around 240°C, which is consistent with references [16] and [19]. The stove temperatures are comprised between 170 and 270°C, which is also consistent with commonly measured temperatures on such wood stove [19]. The wall temperatures behind the stove can rise up to 180°C on the inside (cellular concrete protection) and 60°C on the outside (Figure 6). The flame temperature does not exceed 270°C according to the model, which is clearly under-estimated (typical wood flame temperature is expected to be in the range of 750-1300°C [12] [13] [14] [15]). This kind of under prediction usually arises from a limited spatial resolution [4], but the grid sensitivity analysis performed did not result in higher flame temperatures. The fact that the other temperatures of interest are consistent with what to be expected despite this quite bad flame temperature prediction is somewhat intriguing. This is probably due to the good performance of the flame radiation heat flux formulation coded in FDS, which is always at minimum a given ratio (35% by default) of the local heat release rate in the flame zone, thanks to the inclusion of an empirical radiation loss term, detailed in reference [4].

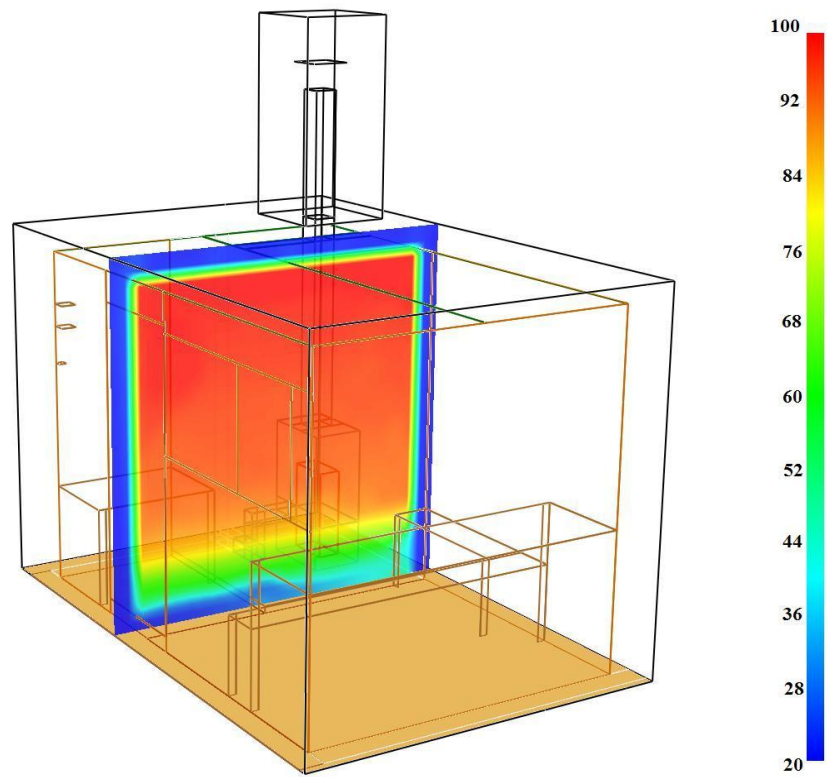

**Figure 5 – Air temperature field in steady-state (°C)**

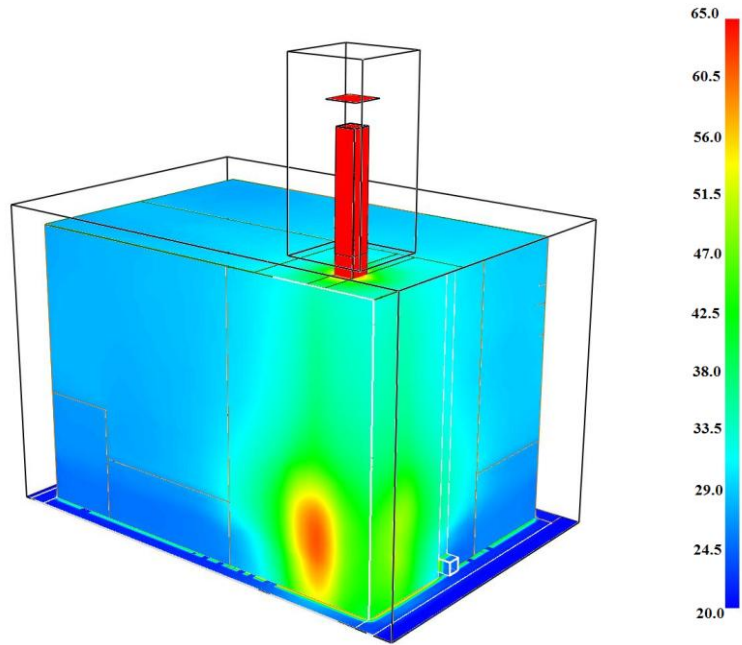

**Figure 6 – Wall temperature field (the hot spot shows the stove location) in steady-state (°C)**

The heat exchange coefficients on the structures are comprised between 2 and 10 W/m²/K (Figure 7). Except for the internal parts of the stove where higher values might have been expected, these results are in very good agreement with the literature for free, mixed and forced convection in air (see for instance references [17] and [18]). The very simple relationships coded in FDS for the heat exchange coefficients [4] thus appear quite reliable here.

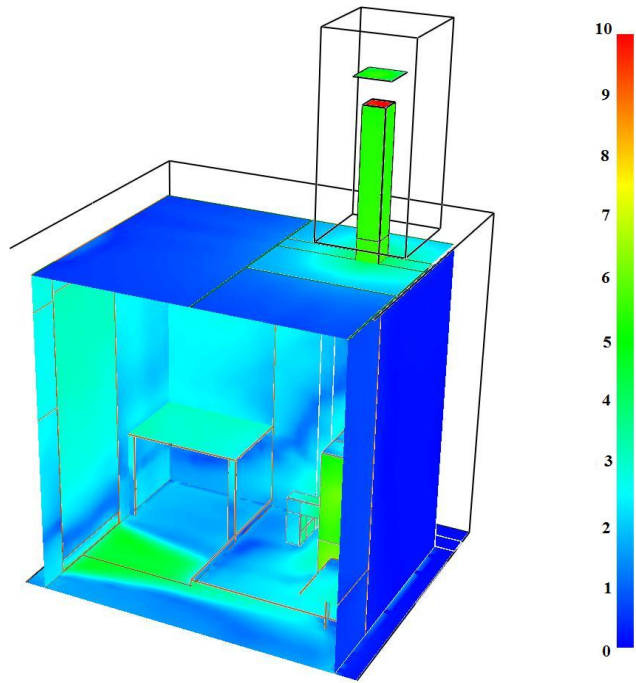

**Figure 7 – Heat exchange coefficient field (W/m²/K)**

A calculation has been performed without the internal part of the chimney, in order to study the impact of its possible rupture, in terms of soot contamination and oxygen concentration. The soot concentration is shown on Figure 8 and the oxygen volume fraction is shown on Figure 9.

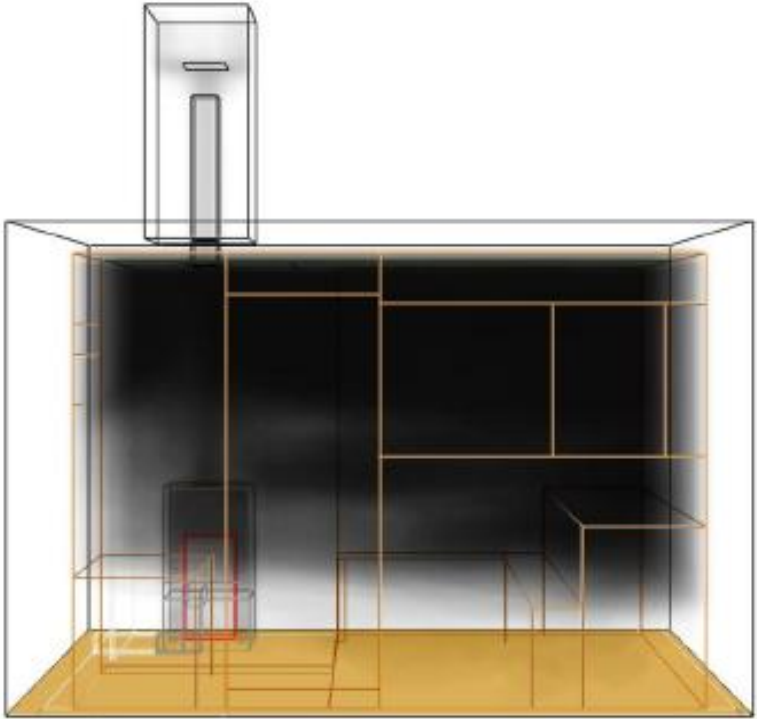

**Figure 8 – Soot concentration (-)**

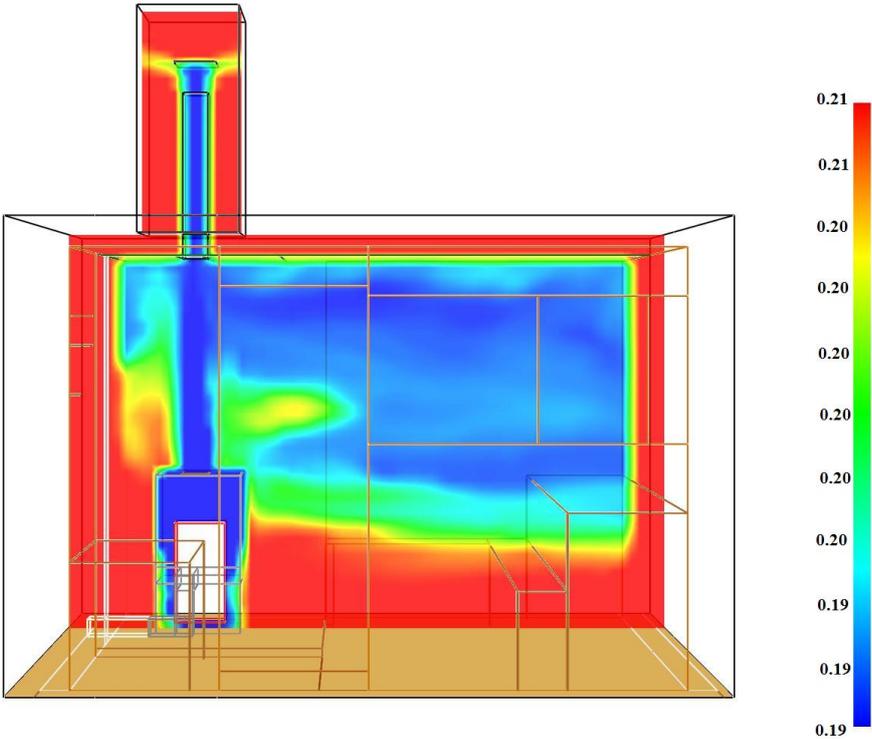

**Figure 9 – Oxygen volume fraction (-)**

The ritual of the sauna implies at some point to pour water on the heated stones of the stove in order to produce a hot steam cloud. In the model, this is done using the nozzle/sprinkler tool, which enables water droplets and water vaporization calculations. Figure 10 shows the water sprinkling on the upper part of the stove (covered with refractory bricks) and Figure 11 shows the steam cloud (there is water vapor in the stove also due to the wood combustion that produces water).

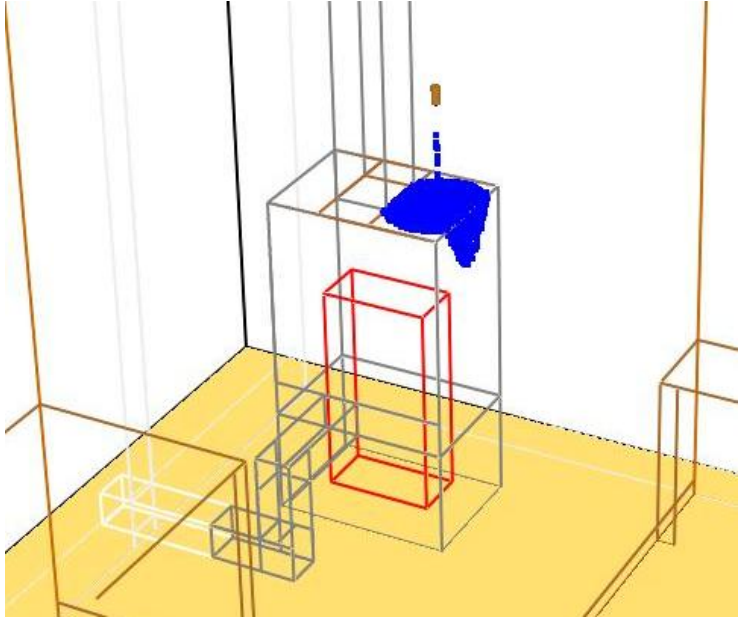

**Figure 10 – Water sprinkling**

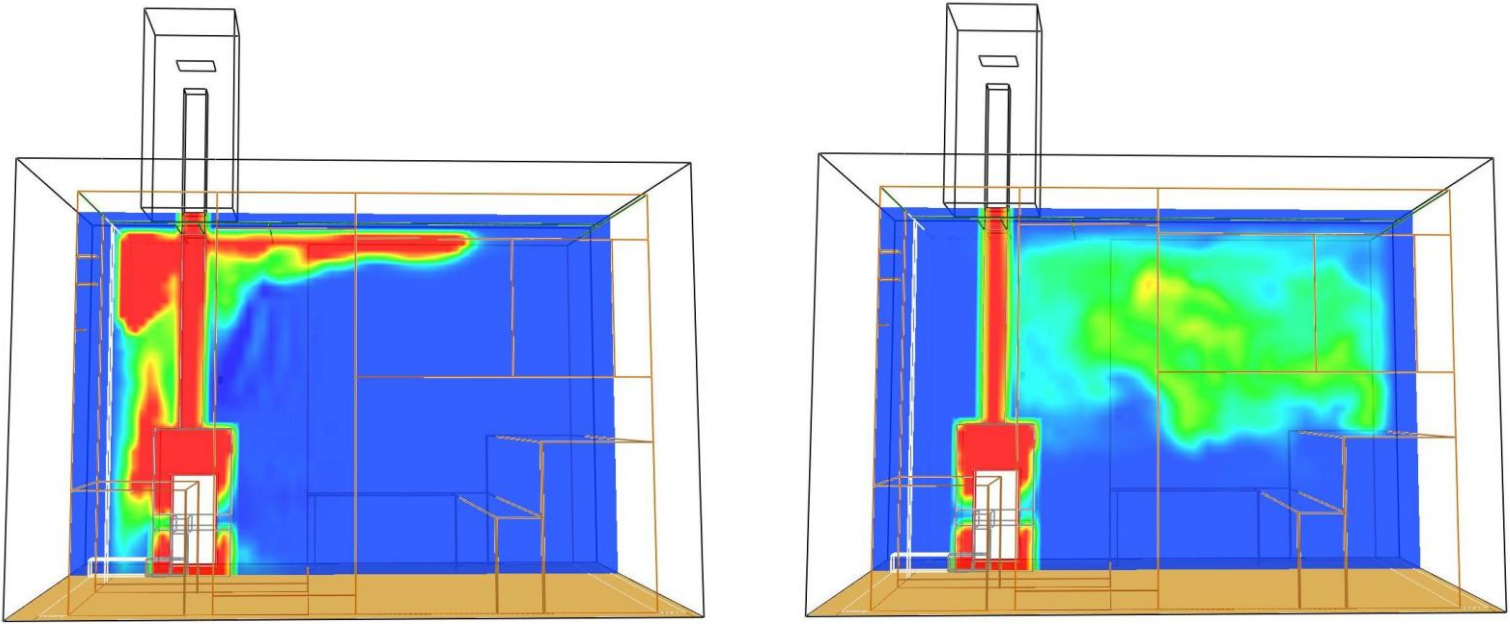

**Figure 11 – Steam cloud volume fraction (-), a few moments after sprinkling on the left, and a few moments later on the right.**

As required for any numerical calculation [3], a grid sensitivity analysis has been performed, by doubling the number of cells in each direction, multiplying by 8 the amount of cells (~1 245 000 cells in total). Temperatures have been compared for the two meshes on several locations (see Figure 12 for one comparison example).

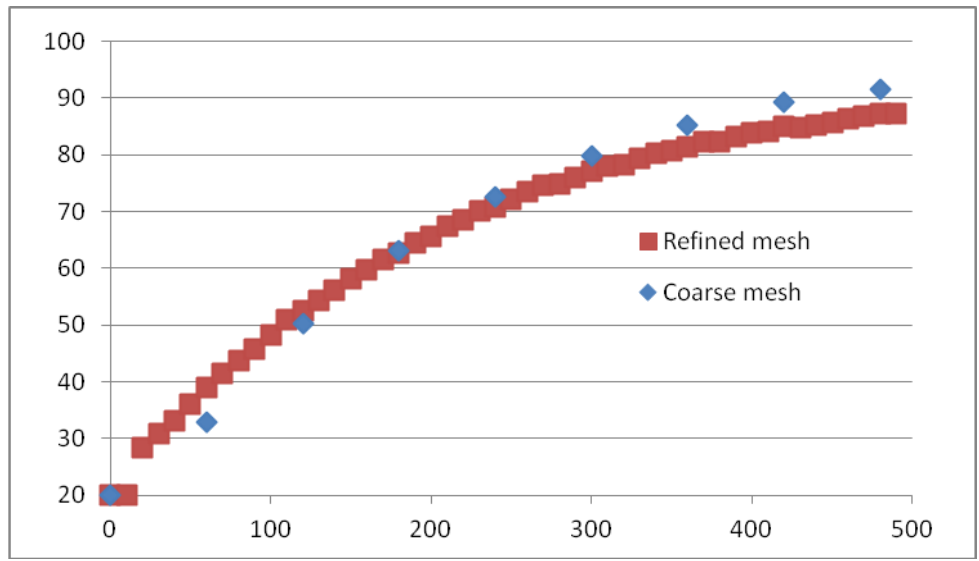

**Figure 12 – Grid sensitivity analysis (air temperature in °C versus time in seconds)**

Although some differences arise from these comparisons, the global solution does not appear to be very sensitive to the grid. The values calculated with the original mesh (~170 000 cells) can hence be considered relevant.

Many other calculations can be (and have been) performed to study the effects of different parameters such as insulation thickness or door opening.

**Conclusion**

This study tends to show that reliable engineering calculations can be performed using the NIST's Fire Dynamics Simulator (FDS) software to estimate the temperatures in a wood-burning stove-heated traditional sauna. This new utilization of FDS (originally a fire safety engineering tool) allows performing Computational Fluid Dynamics (CFD) engineering studies such as insulation thickness estimation, design optimization, safety analysis (oxygen and carbon monoxide levels), and so on.

More detailed studies, compared to precise air and wall temperatures and mass flow rates measurements should be conducted to confirm these preliminary results. The flame temperature prediction should particularly be investigated.